\documentclass[aps,prb,preprint,groupedaddress]{revtex4}

\usepackage{amsmath}
\usepackage{graphicx}
\usepackage{dcolumn}
\usepackage{bm}
\usepackage{subfigure}

\begin{document}

\title{Melting properties of a simple tight-binding \\ 
model of transition metals: I.~The region of half-filled $d$-band}

\author{C. Cazorla$^{1,2,3}$}
\author{D. Alf\`e$^{1,2,3,4}$}
\author{M. J. Gillan$^{1,2,3}$}
\affiliation{
$^{1}$Materials Simulation Laboratory, UCL,
Gower Street, London WC1E 6BT, U.K. \\
$^{2}$London Centre for Nanotechnology, UCL, 
Gordon Street, London WC1H OAH, U.K. \\ 
$^{3}$Department of Physics and Astronomy, UCL, 
Gower Street, London WC1E 6BT, U.K.\\
$^{4}$Department of Earth Sciences, UCL, 
Gower Street, London WC1E 6BT, U.K.}

\begin{abstract}
We present calculations of the free energy, and
hence the melting properties, of a simple
tight-binding model for transition metals in the region of
$d$-band filling near the middle of a $d$-series, the parameters
of the model being designed to mimic molybdenum. The melting
properties are calculated for pressures ranging from ambient
to several Mbar. The model is intended to be the simplest possible
tight-binding representation of the two basic parts of the energy:
first, the pairwise repulsion due to Fermi exclusion; and second,
the $d$-band bonding energy described in terms of an electronic
density of states that depends on structure. In addition to the
number of $d$-electrons, the model contains four parameters, which
are adjusted to fit the pressure dependent $d$-band width
and the zero-temperature pressure-volume relation of Mo. We show
that the resulting model reproduces well the phonon dispersion
relations of Mo in the body-centred-cubic structure, as well as
the radial distribution function of the high-temperature solid and
liquid given by earlier first-principles simulations. Our free-energy
calculations start from the free energy of the liquid and solid
phases of the purely repulsive pair-potential model, without
$d$-band bonding. The free energy of the full tight-binding model
is obtained from this by thermodynamic integration. The resulting
melting properties of the model are quite close to those
given by earlier first-principles work on Mo. An interpretation
of these melting properties is provided by showing how they
are related to those of the purely repulsive model. 
\end{abstract}

\maketitle

\section{Introduction}
\label{sec:introduction}

Many years ago, a combination of experiments, first-principles
calculations and simple models led to the comprehensive
understanding of the low-temperature energetics of transition
metals that we have today.~\cite{pettiforbook} Much more recently, advances in
experimental and first-principles techniques have started to open 
the possibility of achieving the same thing for the high-temperature
phase diagrams, including melting curves, of transition metals over
a wide range of pressures. However, the data obtained so far are
fragmentary and sometimes conflicting,~\cite{errandonea05} and we believe there is now a clear
need to develop simple models analogous to those used to
interpret low temperature data. These models are needed 
in order to elucidate
the fundamental mechanisms that determine 
high-temperature phase diagrams, while
providing a framework within which to interpret and unify experimental
and first-principles data. We describe here 
how a simple parameterised tight-binding
model can be used to calculate the high-temperature free energies
of liquid and solid transition metals, and hence their melting properties,
and we show how the model can help to interpret the available data.
In the present work, we confine ourselves to the case of an
approximately half-filled $d$-band, focusing particularly on the
interpretation of data for molybdenum.

Shock measurements gave the first experimental information about
melting curves at Mbar pressures, and data are available for several
transition metals, including Fe, Mo, Ta and 
W.~\cite{yoo93,brown86,nguyen00,hixson92,mitchell81,holmes04}  
The thermodynamic states accessible in traditional shock experiments
lie on a trajectory called the principal Hugoniot, which provides
only a single point on the melting curve.
On the other hand, major advances in static compression techniques,
based on the diamond anvil cell (DAC), in principle allow entire
melting curves and other phase boundaries to be mapped at pressures and
temperatures up to $\sim 200$~GPa and $\sim 4000$~K. 
Melting data from static techniques
have been reported for Fe, Mo, Ta, W, V and 
Y.~\cite{shen98,jephcoat96,errandonea01, errandonea03,errandonea05} 
There appear to be enormous
differences between the melting curves of some transition metals
from dynamic and static techniques, with the latter giving much
lower melting slopes. The resulting differences of $T_{\rm m}$ at Mbar
pressures can be several thousand K. 

Melting curves from first-principles modelling began to appear over 10
years ago,~\cite{wijs98,alfe99} and there are now several 
well established approaches,
including the calculation of solid and liquid free energies, the
``reference coexistence'' method, and the explicit first-principles
simulation of coexisting solid and 
liquid.~\cite{alfe02a,gillan06,vocadlo02,vocadlo04,
alfe02,alfe04,belonoshko08,alfe03} 
For Fe, all three approaches
have been used, and the agreement between them is 
excellent.~\cite{alfe04b,alfe09} 
Since DFT calculations are parameter-free, and reproduce very 
accurately key quantities such as cold compression curves, phonon
frequencies, Hugoniot curves, and the zero-pressure melting temperatures
of transition metals, there is every reason to expect that their
predictions of melting properties will also be reliable, and there is
considerable evidence that this is the case. For transition metals
for which static and dynamic measurements disagree seriously, 
first-principles calculations support the correctness of the dynamic
measurements.~\cite{cazorla07,taioli07}

Molybdenum is one of the transition metals that have been intensively
studied by DFT simulation, and it illustrates the recent
controversies. Two independent sets of first-principles calculations~\cite{cazorla07,belonoshko08,cazorla08a,cazorla08b,belonoshko04}
agree rather closely with each other and support the high melting
curve deduced from shock measurements, this curve rising far
more steeply with pressure than the flat melting curve obtained from 
DAC data.~\cite{errandonea01} However, the shock measurements~\cite{hixson92} also indicate a solid-solid
phase boundary, which may be the transition interpreted as melting
in the DAC work.~\cite{xiulu08} A similar conflict between high shock and 
first-principles melting curves and a low DAC melting curve is also found in Ta,~\cite{errandonea01,errandonea03,taioli07} 
and it has been proposed that the transition seen in DAC
may also be a solid-solid transition. We believe that simple models may
help to resolve these controversies, by allowing the melting properties
of transition metals to be related to the fundamental mechanisms
that determine their energetics. 

Models for the energetics of transition metals are generally built on
the principle that the total energy can be approximated as the sum
of the electronic band energy and a repulsive pairwise interaction.
The many different models that have been proposed differ mainly in their
representation of the band energy. To explain the broad features
of transition-metal energetics on a scale of several eV,
including the roughly parabolic variations of cohesive energy,
lattice parameter and bulk modulus with band filling, it suffices
to assume a structureless $d$-band density of states (DOS),
whose band width depends only on atomic volume (and chemical element).~\cite{friedel69,pettiforbook}
The simplest total-energy model based on this idea consists of
a sum of repulsive pair potentials plus a position-independent
bonding term depending on the average atomic volume. 
We will refer to this as the REP+VOL model. More sophisticated
types of total-energy models, including the closely 
related second-moment,~\cite{tomanek85}
embedded-atom and Finnis-Sinclair models,~\cite{daw84,finnis84} allow 
the second moment of the local
DOS on each atom to depend on the distances to near neighbours.
However, such models do not contain the physics needed to account for
the well-known low-temperature structural sequence that occurs
through all the transition-metal series, from hexagonal close-packed (hcp),
to body-centred cubic (bcc), to hcp, and finally to face-centred cubic (fcc).
The energy differences of typically a few tenths of an eV 
between these structures are clearly essential for any discussion of
phase diagrams, but they arise from the structure dependent
form of the DOS. 
There are models that account for this by working with higher moments
of the DOS than the second,~\cite{pettiforbook} but a more straightforward approach is to
express the total energy function directly in terms of a 
tight-binding (TB) model.~\cite{goringe97,slater54} In the present work, we use the simplest
possible TB total-energy model, consisting of repulsive pair interactions
plus the sum of single-electron energies calculated from a canonical
$d$-band TB model, without $sp$ bands. We refer to this as the REP+TB model.
With this simple model, we sacrifice the ability
to describe the effect on the DOS of $sp-d$ hybridisation, and the
pressure dependent transfer of electrons between $sp$ and $d$ bands.
We make this sacrifice in order to simplify
the analysis.

The principal question addressed in this paper is: What are the
main parameters that determine the melting curves and other
melting properties of transition metals, and what are the roles
of these parameters? As part of this overall question, we would like
to know at what level of detail we need to describe the
$d$-band bonding. In particular, do we need a detailed description of
the structure-dependent electronic DOS in order to understand melting,
or is a simpler model, such as REP+VOL, sufficient?
In trying to answer these questions, our strategy
will be to relate the melting properties of the REP+TB models to those
of the pure REP model. 

Ultimately, we want to use parameterised tight-binding
models to achieve a systematic overall understanding of the 
melting properties of the entire
family of transition metals. However, even the simple models
used here require rather extensive calculations to treat melting
for a single metal, and for that reason we confine ourselves here
to a narrow range of $d$-band filling in the region of half filling.
We shall present
a simple scheme for fixing the parameters of our model by fitting
to zero-temperature first-principles data, and we shall see that, for
the case of Mo treated here, we reproduce high-temperature
first-principles results reasonably well.

The remainder of the paper is organised as follows. In 
Section~\ref{sec:tbem}, we present our REP+TB model 
for the total energy function,
and we describe the scheme we use to fix the model parameters using
information from $T = 0$~K DFT calculations. In Section~\ref{sec:techniques}, 
we present a variety of tests of the model against DFT, both at $T = 0$~K and
for high-temperature solid and liquid Mo. The procedures used
to calculate the free energies of the pure REP
and REP+TB systems are described in Section~\ref{sec:freenergy}, where we also
report our results for the melting curves and the volume and entropy
of melting. This is followed in Section~\ref{sec:analysis} by an analysis of the
relationships between the melting properties of the REP and REP+TB systems.
Discussion and conclusions are in Section~\ref{sec:discussion}~.

\section{The tight-binding total-energy model}
\label{sec:tbem}

The total energy $U_{\rm tot}$ of our tight-binding (TB) model
for a system of $N$ atoms having position ${\bf r}_i$ is:
\begin{equation}
U_{\rm tot} ( {\bf r}_1, {\bf r}_2 , \ldots {\bf r}_N ) =
\frac{1}{2} \sum_{i \ne j} V_{\rm REP} ( r_{i j} ) +
U_{\rm TB} ( {\bf r}_1 , {\bf r}_2 , \ldots {\bf r}_N ) \; ,
\label{eq:tot}
\end{equation}
where $V_{\rm REP} ( r )$ is a repulsive pair potential and
$r_{i j} = | {\bf r}_i - {\bf r}_j |$. In conventional TB treatments,
the energy $U_{\rm TB} ( {\bf r}_1 , {\bf r}_2 , \ldots {\bf r}_N )$
represents the sum of single-electron energies $\epsilon_n$ of
occupied states, but here we include the effect of thermal
excitation of electrons, so that $U_{\rm TB}$ is actually a
{\em free} energy, defined as:
\begin{equation}
U_{\rm TB} = 2 \sum_n f_n \epsilon_n - T S \; ,
\label{eq:utb}
\end{equation}
where $f_n$ is the Fermi-Dirac occupation number of energy eigenstate
$n$ at temperature $T$, and $S$ is the electronic entropy, given by:
\begin{equation}
S = 2 k_{\rm B} \sum_n \left[
f_n \ln f_n + ( 1 - f_n ) \ln ( 1 - f_n ) \right] \; .
\label{eq:entropy}
\end{equation}
The factors of 2 in Eqns~(\ref{eq:utb}) and (\ref{eq:entropy})
account for spin.
The TB Hamiltonian used to calculate the $\epsilon_n$ is described
next, and the repulsive pair potential is described after that.

\subsection{The canonical $d$-band tight-binding Hamiltonian}
\label{subsec:dband}

Since we include only $d$-electrons in our model, the Hamiltonian matrix
elements $\langle i \alpha | H | j \beta \rangle$ ($i$, $j$ label
atoms, $\alpha$, $\beta$ label atomic orbitals) characterize hopping
transitions of electrons between the d-orbitals $x y$, $y z$,
$z x$, $x^2 - y^2$, $3 z^2 - r^2$ on each atom. We employ an orthogonal
TB model, in which 
$\langle i \alpha | j \beta \rangle = \delta_{i j} \delta_{\alpha \beta}$. The
dependence of the matrix elements on interatomic distance is
taken to be exponential, so that:
\begin{equation}
\langle i \alpha | H | j \beta \rangle =
G_{\alpha \beta} ( {\hat{\bf r}}_{i j} ) A_b \exp ( - r_{i j} / R_b ) \; .
\end{equation}
The factor $G_{\alpha \beta}$ depends on the unit vector
${\hat{\bf r}}_{i j} = ( {\bf r}_i - {\bf r}_j ) / r_{i j}$ in the
direction from ${\bf r}_i$ to ${\bf r}_j$, and it is well
known that it can be expressed in terms of three basic matrix
elements dd$\sigma$, dd$\pi$ and dd$\delta$. Here, we assume the canonical
ratios~\cite{andersen73}
dd$\sigma$ : dd$\pi$ : dd$\delta$ = $-6$ : $4$ : $-1$. For convenience,
and without loss of generality, we assume the diagonal elements
$\langle i \alpha | H | i \alpha \rangle$ to be zero. In order to
simplify the numerical simulations, we cut off the matrix elements
so that they vanish beyond a distance $R_{\rm cut}$. The exponential
is replaced by a cubic polynomial in the interval $R_1 < r < R_{\rm cut}$~,
the polynomial coefficients being chosen to ensure continuity of
$\langle i \alpha | H | j \beta \rangle$ and its first derivative 
at $R_1$ and $R_{\rm cut}$. For the Mo model developed here,
we chose $R_1 = 4.7$~\AA\ and $R_{\rm cut} = 4.9$~\AA.

The TB density of states (DOS) $n_d ( E )$, defined as:
\begin{equation}
n_d ( E ) = \frac{2}{N} \sum_n \delta ( E - \epsilon_n ) \; ,
\end{equation}
is normalized so that $\int n_d ( E ) \, d E = 10$. Since the
trace of $\langle i \alpha | H | j \beta \rangle$ is zero, the
first moment $\mu_d^{(1)}$ of the DOS, defined as:
\begin{equation}
\mu_d^{(1)} = \int E n_d ( E ) \, d E \left/
\int n_d ( E ) \, d E \right.
\label{eqn:1stmoment}
\end{equation}
is zero. To fix the values of $A_b$ and $R_b$, we require that the second
moment $\mu_d^{(2)}$ of the DOS of our model, defined as:
\begin{equation}
\mu_d^{(2)} = \int E^2 n_d ( E ) \, d E \left/
\int n_d ( E ) \, d E \right. \; ,
\label{eqn:2ndmoment}
\end{equation}
should agree with the volume dependent $d$-band second moment given by DFT.

To apply this procedure to bcc Mo, we 
performed DFT calculations using 
the full-potential linearized augmented plane-wave method 
(FP-LAPW)~\cite{andersen75,koeling75,koeling77,singh94} as 
implemented in the WIEN2k code.~\cite{blaha01} 
We used the Wu-Cohen~\cite{wu06} form of generalized gradient 
approximation (GGA), which is known to perform well for transition 
metals.~\cite{tran07,cazorla08}
Local orbitals are added to the standard LAPW basis
in order to describe valence and semicore states. The technical parameters
in the calculations were set as in Ref.~[\onlinecite{cazorla08}]~.
The total and projected $d$-channel densities of states were obtained 
by using the modified tetrahedron method of 
Bl\"ochl {\emph et al.}~\cite{blochl94}, and for the projection
we used an atomic sphere radius of typically $1.32$~\AA.
We found that $A_{b} = 18.5745$~eV and $R_{b} = 0.8950$~\AA\ give very 
good agreement
with the DFT results for $\mu_d^{(2)}$ at $P = 0$ 
and $350$~GPa (see Table~I) and these values 
are used throughout this work.

The quantity $\mu^{(2)}_{d}$ is closely related to the 
width of the $d$-band $W_{d}$, 
which is the difference between the lowest and highest energy levels, 
$E_{d}^{b}$ and $E_{d}^{t}$ respectively, in the $d$-band DOS. 
In DFT calculations, the bottom of the $d$-band $E_{d}^{b}$ 
can be determined by direct 
inspection of the DOS, whereas $E_{d}^{t}$ may be difficult to identify  
because of hybridization of states with different angular momenta 
(see Fig.~\ref{fig:dos0T}). Here we identify $E_{d}^{t}$ with 
an abrupt drop in the projected
$d$-DOS at high energies followed by a smooth continuum. 
For Mo, we find that at equilibrium $E_{d}^{b}$ and $E_{d}^{t}$ 
are $-5.5$ and $4.6$~eV, respectively, while at a pressure of $P = 350$~GPa 
they are $E_{d}^{b} = -10.8$ and 
$E_{d}^{t} = 8.6$~eV~(see Fig.~\ref{fig:dos0T}). 
As shown in Table~I, these values compare well with the tight-binding 
results obtained with 
the $A_{b}$ and $R_{b}$ values quoted above.

In order to reproduce the energy difference between the Fermi 
level and bottom of the
$d$-band and the form of the $d$-DOS near $E_{F}$, 
we treat the number of $d$ electrons $N_{d}$ as an 
adjustable parameter.~\cite{paxton96} 
This is important, since many properties of transition 
metals are understood in terms  
of the form of the electronic DOS near the Fermi 
level (e.g. the relative stability of different structures, 
electronic specific heat, etc). 
For Mo, we find that $N_{d} = 4.3$, rather than $N_{d} = 5.0$, 
reproduces quite well the 
DFT results over a range of pressures (see Fig.~\ref{fig:dos0T}). 
We use this value of $N_d$, unless stated otherwise.

\subsection{The repulsive pair potential}
\label{subsec:repulsive}

The pair potential $V_{\rm REP} ( r )$ is also assumed to have an
exponential form:
\begin{equation}
V_{\rm REP} (r) = A_r \exp ( - r / R_r ) \; .
\label{eq:repulpot}
\end{equation}
The parameters
$A_r$ and $R_r$ are chosen so as to reproduce as closely as 
possible the measured $P$-$V$ curve
of bcc Mo at low temperatures. This is essentially the same as fitting
to DFT, since with the Wu-Cohen functional the DFT and experimental
$P$-$V$ curves are almost identical.  
The values $A_r = 3164.3454$~eV~ and $R_r = 0.3350$~\AA\ give  
excellent agreement with experimental data 
of Ref.~[\onlinecite{hixson92}],
and DFT calculations (Fig.~\ref{fig:eos}), and we use them throughout 
this work. The same spatial cut-off distance and smoothing as used for 
the Hamiltonian matrix elements is applied to the repulsive pair potential.

\section{Simulation techniques and tests of the model}
\label{sec:techniques}

\subsection{Molecular dynamics simulation}
\label{subsec:mds}

All the calculations on our TB model were performed with the
OXON code,~\cite{oxona,oxonb,oxonc,oxond} using diagonalization
of the Hamiltonian for each set of ionic positions. In the
molecular dynamics (m.d.) simulations, we used the
Verlet algorithm to integrate Newton's equations of motion,
with a typical time step of $1.25$~fs.
The total force acting on each atom is the exact derivative  
of the total energy $U_{\rm tot}$ with respect to its atomic position.
Our m.d. simulations were performed in the canonical NVT ensemble, using
Andersen's thermostat to avoid errors
due to lack of ergodicity.~\cite{andersen80} In using this thermostat,
the atomic velocities were randomized by drawing
them from a Maxwellian distribution every $0.2$~ps.
A typical m.d. run consisted of $2$~ps for equilibration, followed by
$10$~ps for the calculation of averages. The m.d. simulations
were performed on a $6\times 6 \times 6$ supercell containing $N = 128$
atoms, and $\Gamma$-point sampling was used to integrate 
over the first Brillouin zone. 
Pressure was obtained directly in each run using the virial formula.

\subsection{Tests of the model}
\label{subsec:tests}

We have performed a series of zero and finite-temperature tests  
of our model in order to assess its accuracy compared with
first-principles results and experimental data. 
In our first test, we evaluated the relative stability of the different
crystal structures at different volumes. To this end, we 
computed the energy differences $\Delta E$ of the hcp and fcc 
structures with respect to the bcc structure at equilibrium 
($V_{0} = 15.55$~\AA$^{3}$/atom) and a pressure of 
$P = 350$~GPa ($V = 9.50$~\AA$^{3}$/atom)~. At equilibrium, we found
fcc-bcc and hcp-bcc differences of 0.40 and 0.46~eV with DFT, compared
with 0.26 and 0.42 with TB.
We thus reproduce correctly the relative zero-pressure stability 
of the different crystal structures, 
though we predict the fcc phase to be appreciably more stable than hcp.
At $P = 350$~GPa, we found fcc-bcc and hcp-bcc differences 
of 0.30 and 0.32~eV with DFT, compared with 0.42 and 0.30~eV with TB.
At this pressure, we thus
predict that the energy of the hcp phase is lower than 
that of the fcc phase, but bcc is correctly predicted to be the
most stable structure. We note that at both pressures 
the agreement between the values of 
$\Delta E_{\rm hcp-bcc}$ obtained with DFT and TB is very good.

We have also computed the phonon frequencies of bcc Mo 
at the experimental equilibrium volume 
$V_{0} = 15.55$~\AA$^3$/atom~.~\cite{hixson92}
Our calculations are based on the small-displacement 
method,~\cite{kresse95,dario98} 
and we have used a supercell containing $64$ atoms 
and $16\times16\times16$ ${\bf k}$-point 
grid over the first Brillouin zone. 
In Fig.~\ref{fig:phonons}~, we show our results together with experimental
data from Ref.~[\onlinecite{hixson92}] and \emph{ab initio} calculations from 
Ref.~[\onlinecite{cazorla07}] for comparison; 
the agreement between the TB curves and the others 
is unexpectedly good, given the simplicity of our model.  
We note that the experimental phonon anomaly near the 
H point $( 1 , 0 , 0)$ is not well reproduced by
either TB or DFT.~\cite{cazorla07,souvatzis08} 
We have calculated the phonon frequencies also for the
fcc and hcp structures of our TB model. We find that
for $N_{d} = 4.3$ and $N_{d} = 5.0$ there are always imaginary frequencies, 
so that these structures are unstable, at least at $T = 0$. 
It worth noting that for slightly larger
$N_{d}$ values the fcc and hcp 
phases become stable at high pressures; 
for instance, for $N_{d} = 5.2$ fcc becomes stable at  
$P \simeq 400$~GPa. 

The finite-temperature tests of our model
include an analysis of the structure of the solid and liquid 
at different pressures.
In Fig.~\ref{fig:grtest}~, we plot the radial distribution 
function obtained from long (total simulation time $\sim 10$~ps) 
DFT and TB m.d. runs. 
The solid phase is simulated at $T = 2000$~K and 
$P = 50$~GPa~, while the liquid is at $T = 8250$~K and $P = 250$~GPa.
(These are states well below and above the melting curve
of Mo given by first-principles calculations.)~\cite{cazorla07,belonoshko08}
In both phases, the DFT and TB curves agree very well, the main difference
being that TB gives interatomic distances slightly smaller
than those from DFT simulations. 

In Fig.~\ref{fig:dostest}~, we show the electronic DOS 
of solid and liquid Mo obtained at the same thermodynamic conditions 
as for the radial distribution function. Although 
the DOS's obtained with TB and DFT are not identical,
the corresponding band-widths and energy differences
$E_{F} - E_{d}^{b}$ are very similar, especially for the 
crystal.     
 
The main conclusion from all these tests is 
that, in spite of the formal simplicity of our TB model, it 
reproduces quite reliably many important properties of solid and liquid Mo.

\section{Free energy and melting properties of the model}
\label{sec:freenergy}

Our overall strategy to obtain the melting properties of our 
model is based on the calculation of the Helmholtz free energy 
$F_{\rm tot} ( V , T )$ of the solid and liquid phases. 
To obtain $F_{\rm tot} ( V , T )$~, we start from the Helmholtz free energy 
$F_{\rm REP} ( V , T )$ of the purely repulsive system described by the 
pair potential $V_{\rm REP} (r)$~, and use thermodynamic integration
to determine the difference $F_{\rm tot} ( V , T ) - F_{\rm REP} ( V , T )$
at fixed $( V , T )$. This thermodynamic integration
is based on the general principle that for total-energy functions
$U_{0}$ and $U_{1}$~, the difference of the corresponding free energies
$F_{0} ( V , T )$ and $F_{1} ( V , T )$ at state point $( V , T )$ is given by
\begin{equation}
F_{1} - F_{0} = \int_{0}^{1} \langle \Delta U \rangle_{\lambda}~d\lambda~,
\label{eq:freediff}
\end{equation}
where $\Delta U = U_{1} - U_{0}$ and  $\langle \cdot \rangle_{\lambda}$ 
denotes the thermal average in the ensemble governed by the total 
energy function $U_{\lambda} = ( 1 - \lambda ) U_{0} + \lambda U_{1}$~.
In practice, we use this type of thermodynamic integration to determine
$F_{\rm tot} ( V , T ) - F_{\rm REP} ( V , T )$ at a set of $( V , T )$
states, the value of $F_{\rm tot} ( V , T )$ at other states being obtained
by integrating the relations 
$P = - ( \partial F_{\rm tot} / \partial V )_{T}$ and
$E_{\rm tot} = ( \partial ( \beta F_{\rm tot} ) / \partial \beta )_{V}$,
where $\beta = 1 / k_{B} T$ and $E_{\rm tot}$ is the total 
internal energy of the system. 
The starting point of all the calculations is
the free energy $F_{\rm REP} ( V , T )$ of the pure exponential system.
Surprisingly, the thermodynamic properties of this system appear
not to have been studied before, so we
have performed our own calculations of $F_{\rm REP} ( V , T )$,
as described next. 

\subsection{Free energy and phase diagram of the pure exponential model}
\label{subsec:pdexp}

Details of the calculations of the free energy of the pure exponential model 
will be reported elsewhere, and here we give only a brief summary.
Our values of $F_{\rm REP} ( V , T )$ 
were obtained by thermodynamic integration  
(Eq.~(\ref{eq:freediff})), using as reference system the inverse-6 
system interacting with pair potential
$V_{\rm inv6} (r) = A / r^{6}$. We take the Helmholtz free energy
of this system from Ref.~[\onlinecite{laird92}] for the liquid, bcc
and fcc phases, and from Ref.~[\onlinecite{prestipino05}] for the 
hcp phase. The thermodynamic integration calculations
were performed at a series of $( V , T )$ points in which 
the free-energy difference 
$F_{\rm REP} - F_{\rm inv6}$ was calculated by averaging 
$V_{\rm REP} - V_{\rm inv6}$ over long molecular 
dynamics runs in which $U_{\lambda}$ (see Eq.~(\ref{eq:freediff})) 
was varied continuously 
at a switching rate that guaranteed reversibility (that is, adiabatically).
For the solid phase, we considered $12$ volumes distributed uniformly over
the interval $9.68 \leq V \leq 30.80$~\AA$^{3}$/atom, and the 
temperature was set to $T = 1000$~K in all cases.
For the liquid phase, $15$ points within the same volume 
range as used for the solid
and temperatures taken at intervals of $1000$~K from 
initial guessed melting temperatures 
up to $10000$~K~, were considered.
We determined $F_{\rm tot} ( V , T )$ at the other state points 
by performing thermodynamic
integration with respect to pressure and internal energy.

Our calculated $F_{\rm REP} ( V , T )$ values were cross-checked against 
simulations in which the liquid coexists with the bcc, fcc or hcp solid. 
These coexistence simulations were performed in the $ ( N , V, T ) $ ensemble,
and we used the techniques explained in 
Refs.~[\onlinecite{cazorla07}] and 
[\onlinecite{vocadlo04}]~. Simulation boxes containing 
up to $10,000$ atoms were used 
in the calculations. For each pair of coexisting phases, 
the pressure dependence
of the melting temperature $T_{\rm m} (P)$ was fitted to the equation
\begin{equation}
T_{\rm m} ( P ) = a \bigg[ \left( 1 + \frac{P}{b} \right)^{c} - 1 \bigg]~,
\label{eq:simonvar}
\end{equation}
which resembles the so-called Simon equation~\cite{simon29}, but is adjusted
to ensure that $T_{\rm m} = 0$ at $P = 0$. 
Results for the bcc, hcp and fcc melting curves are 
shown in Fig.~\ref{fig:meltingexp}~. 
For the bcc melting curve, the values are accurately reproduced with
parameters $a = 564.6$~K, $b = 1.69$~GPa and $c = 0.5236$.
The volumes and enthalpies per atom of the coexisting solid
and liquid were obtained from independent molecular dynamics simulations
performed on supercells containing $1,000$ atoms at $( P , T )$ points 
on the melting curve. 
The melting volumes and entropy of fusion for the bcc melting curve
are given in Table~II.
In fact, the melting curves obtained from the coexistence simulations
were not perfectly consistent with the Helmholtz free energy results. 
We have searched carefully  for the source of these errors, and we think
it is possible that they may come from small imprecisions of the free
energies of the inverse-6 system.  
To correct for these errors, we shifted 
the free energies of the bcc, hcp and fcc phases with respect 
to that of the liquid;
the corrections depend solely on temperature, and are typically 
$10-20$~meV/atom. We note that differences between energy shifts of all 
three crystal structures amount to less than $5$~meV/atom, 
so therefore, total free energy differences between the bcc, fcc, and hcp phases,
or equivalently their relative stability, are not affected appreciably 
by our corrections.

As a further cross-check, we calculated the free energies of
the bcc and fcc phases by thermodynamic integration, starting from
a different reference system. For this, we took a harmonically
vibrating solid, with the harmonic force-constant matrix
calculated for a system of particles interacting via the repulsive 
pure exponential pair potential at volume $V = 14.19$~\AA$^{3}$/atom ($P \sim 204$~GPa). 
These calculations are based on the small-displacement method.~\cite{kresse95,dario98} 
For both the bcc and the fcc phases,
the small discrepancies between the free energies obtained from
the inverse-6 and harmonic reference systems are typically 
$10 - 20$~meV/atom at temperatures near melting.

\subsection{Free energy and melting properties of the TB model}
\label{subsec:melting}

Our thermodynamic integration calculations to determine
the difference $F_{\rm tot} - F_{\rm REP}$ were performed 
by varying $\lambda$ adiabatically from $0$ to $1$ over 
a time of $9$~ps. Simulation boxes containing $128$ atoms
and $\Gamma$-point sampling over the first 
Brillouin zone were used.
The quantity $\Delta U$ in these calculations
(see Eq.~(\ref{eq:freediff})) is the TB band free-energy $U_{\rm TB}$
(see Eqs.~(\ref{eq:utb}) and (\ref{eq:entropy})).  
In order to reduce errors due to
non-adiabaticity, we perform a complete cycle in which $\lambda$
goes from $0$ to $1$ and back again, and to reduce statistical
errors this whole cycle is repeated. The thermodynamic integral
$\int_0^1 d \lambda \, \langle U_{\rm TB} \rangle_\lambda$ is
obtained as the average of the values in the four half-cycles.
The typical standard deviation of these values is less
than $10$~meV/atom~.  
An example of the $\langle U_{\rm TB} \rangle_{\lambda}$ values obtained
over a whole run at $V = 15.55$~\AA$^{3}$/atom~
and temperature $T = 5048$~K is shown in Fig.~\ref{fig:adiabatic}.
We note that $\langle U_{\rm TB} \rangle_{\lambda}$
varies typically by $\sim 0.5$~eV/atom as $\lambda$ varies between
$0$ and $1$~.

These thermodynamic integration calculations were performed
at ten $( V , T )$ states in each of the liquid and solid phases.
For the solid phase, a temperature of $ T = 1000 $~K was
chosen for all the volumes; volumes were drawn uniformly from the interval 
$9.68$ - $16.32$~\AA$^{3}$/atom.
For the liquid phase, temperatures of typically $3000$~K
above the melting curve of the repulsive potential
were chosen and the same set of volumes as for the solid was used.
The value of $F_{\rm tot} ( V , T )$ at the other thermodynamic states
was obtained by thermodynamic integration with respect to pressure
and internal energy.

In Fig.~\ref{fig:melting}~, we report the melting line of our 
TB model for $d$-band fillings $N_{d} = 4.3$ and $5.0$,
obtained from the Helmholtz free energy calculations described above. 
We have considered different $N_d$ values in order
to assess the effect of this on the melting properties. 
Since our harmonic calculations showed that only the bcc 
structure is vibrationally
stable, we will report only results for melting from the bcc structure.
In practice, once the free energies $F_{\rm tot} ( V , T )$ 
of the liquid and solid phases are known,
we have determined the melting pressure $P_{\rm m}$ and volumes 
of the liquid and solid phases 
at each temperature by the Maxwell double-tangent construction. 
The Simon formula $T_{\rm m} = a \left( 1 + P_{\rm m}/b \right)^{c}$ 
was then used to fit 
our results and interpolate at any desired pressure. For 
$N_{d} = 4.3(5.0)$, the values of the Simon parameters are
$a = 2865.9 (1678.6)$~K, 
$b = 118.3 (35.1)$~GPa and $c = 0.8530(0.6376)$. 
In Table~III, we report results for the fractional 
change of volume $\Delta V / V_{s}$ and entropy
of fusion $\Delta S$ at points on the melting curves.

In Fig.~\ref{fig:melting}, we also plot the melting curves 
of the pure exponential system and of Mo obtained from 
DFT calculations.~\cite{cazorla07}
Although accurate reproduction of real-world data is not 
the main objective of this work,
we note that our model (case $N_{d} = 4.3$) gives 
very good agreement with  
the $P = 0$ melting temperatures 
for Mo of $T_{\rm m} = 2883$~K from experiment~\cite{shaner77}, 
and $T_{\rm m} = 2894$~K from DFT simulations.~\cite{cazorla07}
With increasing $P$, significant discrepancies between 
the TB ($N_{d} = 4.3$) 
and DFT melting curves appear. However, good 
agreement is partly recovered at high-$P$ and 
high-$T$ for $N_{d} = 5.0$. 

\section{Analysis of melting relationships}
\label{sec:analysis}

We pointed out in the Introduction that the gross features of transition
metal energetics at $T = 0$~K can be understood on the basis of a model
in which the structure of the electronic density of states DOS is ignored.
This suggests that the simplest possible model for understanding 
the melting behaviour
of transition metals is to add to the free energy 
of the pure exponential 
model $F_{\rm REP} ( V , T )$ a bonding term $E_{d} (V)$ that depends only on 
volume and does not depend on temperature or on the phase of the system. 
To test this idea, we have carried out numerical calculations in which we have
set $E_{d} ( V )$ equal to the bonding energy contribution to
the total energy $U_{\rm tot}$ of the bcc solid at zero 
temperature ($N_{d} = 4.3$).
As expected,  $E_{d} (V)$ varies between $-10$ and $-20$~eV/atom 
over the volume
range of interest. The resulting melting curve is shown in 
Fig.~\ref{fig:meltingexpvol}~.
This very simple model necessarily shifts the melting 
curve upwards, and our results
show that the computed melting temperatures are seriously overestimated, 
typically by around $50$\%~. 
This result shows that there must be a significant dependence 
of the bonding energy
on structure for given volume in the region of the melting curve. To illustrate
this, we show in Fig.~\ref{fig:deltaf} (Top) the 
bonding free energy $\Delta F = F_{\rm tot} - F_{\rm REP}$
as a function of volume at $ T = 6000$~K for the liquid 
and bcc solid. Remarkably,
the difference between $\Delta F$ for liquid and solid is rather constant
and has a value of $\sim 0.2$~eV/atom, $\Delta F$ being lower in the liquid.
This means that the structure dependence of the bonding stabilizes 
the liquid phase over the solid and therefore 
lowers the melting curve.

It is interesting to ask whether $\Delta F$ is significantly influenced 
by the response of the structure to the presence of the tight-binding energy.
To answer this, we show in Fig.~\ref{fig:deltaf} (Bottom) the
quantity $\Delta F - \langle U_{\rm TB} \rangle_{\rm REP}$ 
where $\langle U_{\rm TB} \rangle_{\rm REP}$
is the thermal average of $U_{\rm TB}$ evaluated in the ensemble of 
the $V_{\rm REP}$ potential. The results show that 
$\Delta F - \langle U_{\rm TB} \rangle_{\rm REP}$
is quite significant in both phases. Moreover, the 
difference of this quantity for the 
liquid and solid indicates that the structure of 
the liquid responds significantly more
than the solid to the presence of the TB energy. This 
effect contributes significantly 
to the lowering of the melting curve. 
 
\section{Discussion and conclusions}
\label{sec:discussion}

The present work is intended as a step towards developing
an overall understanding of the phase diagrams of entire transition-metal
series over a wide range of pressures and temperatures. At $T = 0$~K,
generalized phase diagrams (GPD) as a function of pressure $P$ and
atomic number $Z$ can be computed by DFT, and we recently
reported a phase diagram of this kind for the 4d 
series.~\cite{cazorla08} The construction of
a complete GPD as a function of $P$, $T$ and $Z$ using DFT is too
difficult at present, but we believe that it should be feasible using
TB models of the kind described here. With this in mind, it is
encouraging that our REP+TB model for Mo, parameterized using only
$T = 0$~K data, reproduces quite well the melting curve and properties
of the high-$T$ solid and liquid known from DFT. We have used the
same REP+TB model, parameterized using the same scheme, for most of
the other 4d metals, and we hope to report $P$-$T$ phase diagrams
for them in due course. We note that corresponding-states arguments
will allow the free energies and melting data for the pure REP
model reported here to be used to obtain the free energies of
all these transition metals by thermodynamic integration.

Since the properties of Mo over a wide range of $P$ and $T$
seem to be quite well described by our REP+TB model, it is
natural to ask how the melting properties of the model are related
to those of the pure REP model, consisting only of exponential repulsion.
The melting temperature of REP goes to zero as $P \rightarrow 0$,
so it is clear that the TB energy is crucial in determining the
$T_{\rm m}$ of transition metals at ambient $P$. However, we might
expect the repulsion to be increasingly dominant at high $P$.
There is an interesting connection here with the melting properties
of the Lennard-Jones (LJ) model for rare gases. It was recognised
long ago~\cite{hansen70} that as $P \rightarrow \infty$, the attractive
$r^{-6}$ potential of LJ has diminishing influence on the properties
of the coexisting solid and liquid, so that the volume and entropy
of fusion tend to those the soft-sphere repulsion $r^{-12}$ model.
As we have shown, the melting curves of our REP+TB model for Mo with
$N_d = 4.3$ and $5.0$ do become close to that of pure REP at high $P$.
Furthermore, the relative melting volumes $\Delta V / V_s$ of
REP+TB for both $N_d$ values decrease steadily with increasing $P$,
in a way that is consistent with convergence towards the
melting volume of pure REP. However, this convergence is slow,
since even at $P \simeq 400$~GPa,
$\Delta V / V_s$ for REP is 1.1, while for REP+TB it is $\sim 2.1$
and 1.5 for $N_d = 4.3$ and 5.0 respectively. The entropy of
fusion $\Delta S$ is 0.74~$k_{\rm B}$ for REP over the whole
pressure range studied. For $N_d = 4.3$, $\Delta S$ decreases
steadily towards this value with increasing $P$, while for
$N_d = 5.0$ it remains a little above this value for all $P$.

In the Introduction, we asked what are the main parameters that
determine the melting properties of transition metals. The success
of our REP+TB model for Mo suggests that the two parameters
$A_r$ and $R_r$ specifying the strength and range of the 
interatomic repulsion, the strength and range $A_b$ and $R_b$ of
the TB matrix elements, and the number $N_d$ of $d$-electrons,
may be enough. (Firm conclusions must, of course, await TB
calculations on other transition metals.) Our simulations
show clearly that a description of the volume-dependent $d$-band width
by itself is not enough. The very large d-bonding energy
$E_d ( V )$ is described by the very simple REP+VOL model, but we
have shown that this always raises the melting curve well above that
of REP, and gives $T_{\rm m} ( P )$ predictions that agree poorly
with the actual melting curves of REP+TB. The melting curves
are substantially reduced below those of REP+VOL by the rather
small shifts of relative free energies of solid and liquid included in
the full REP+TB model. We have seen that a significant contribution
to these shifts comes from the response of the system (particularly
the liquid) to the presence of the TB energy.

The present work may shed light on recent interpretations of the
flat melting curves inferred from DAC measurements. It has been
proposed that the directional bonding associated with partially
filled $d$-bands may give rise to ``preferred local structures''
having icosahedral short-range order in the liquid 
phase.~\cite{ross07a,ross07b} It was suggested that
the formation of these local structures lowers the free energy of
the liquid, and hence depresses $T_{\rm m}$. By contrast with
simpler models, such as the embedded-atom model, the TB model
we use fully includes directional d-bonding, and our simulated
liquid would presumably exhibit the effects of ``preferred local
structures'', if they were present. The same can be said of
the DFT simulations that have been reported on Mo. Nevertheless,
both the present TB calculations and the earlier DFT simulations
give much steeper melting curves than the DAC measurements, and
this indicates that the full inclusion of directional d-bonding
does not lead to low melting curves, in contradiction with
the suggestions of Refs.~[\onlinecite{ross07a,ross07b}]~.

Our TB model represents only the $d$-band, and ignores the $s-p$ band.
This means that, although it mimics the pressure-dependent width
of the $d$-band and gives the main features of the DOS, it cannot 
reproduce the fine details,
since it neglects hybridization of d-states with sp-states. It also
means that the number of $d$-electrons $N_d$ has to be treated as
an adjustable parameter, and we do not include the dependence
of this number on pressure or structure. It has been 
suggested~\cite{ross04} that the dependence of $N_d$ on structure might lead to the very
flat melting curves inferred from DAC experiments. However, these ideas
are not supported by DFT simulations, which fully include
structure-dependent sp-d transfer, but nevertheless give melting
curves that rise much more steeply than those from DAC. The fact
that the present $d$-band-only TB models give melting curves in
reasonable agreement with DFT confirms that sp-d transfer is
not expected to give flat melting curves.

A possible resolution of the conflict between shock and first-principles
melting curves on one side and DAC melting curves on the other side has
emerged recently, at least for some transition 
metals.~\cite{belonoshko08,cazorla08a,belonoshko08b}
DFT simulations of Mo have shown that, although bcc is 
the most stable structure at low $T$
up to over 600~GPa, another structure, perhaps fcc or hcp, is likely to
become more stable than bcc at much lower $P$ and temperatures well
below the melting curve. The suggestion is that the transition
interpreted as melting in DAC experiments on Mo may actually be
the transition between bcc and this other structure. 

Our main conclusions are as follows: A simple tight-binding
model, parameterized using data for the volume-dependent
$d$-band width and the cold compression curve of Mo reproduces
reasonably well the melting curve and the properties of high-$P$/high-$T$
solid and liquid Mo known from DFT simulations; the model allows
us to analyse the physical mechanisms that determine the melting
properties, and to assess suggested explanations for
the anomalously low melting curves inferred from static
compression experiments. We hope to report soon on TB
calculations of melting properties across the whole 4d series.

\acknowledgments
The work was supported by EPSRC-GB Grant No. EP/C534360, which
is 50\% funded by DSTL(MOD). The work was conducted as
part of a EURYI scheme award to DA as provided by
EPSRC-GB (see www.esf.org/euryi).

\clearpage

\begin{table}
\begin{center}
\begin{tabular}{c  c c c }
\hline
\hline
\qquad  & $\qquad W_{d} \qquad$ & $\qquad \mu^{(2)}_{d} \qquad$ & $\qquad E_{F} - E_{d}^{b} \qquad$\\
\hline
${\rm DFT}$ & $ 10.1 (19.4)$ & $ 7.22 (23.66) $ & $5.88 (10.78) $ \\
${\rm TB}$ &  $ 10.8 (19.4)$ & $ 7.29 (21.25) $ & $5.85 (10.86) $ \\
\hline
\hline
\end{tabular}
\end{center}
\caption{Calculated $d$-band width
$W_{d} = E_{d}^{t} - E_{d}^{b}$, second moment $\mu^{(2)}_{d}$
and energy difference $E_{F} - E_{d}^{b}$ from DFT and TB
at $P =0$~GPa ($P = 350$~GPa in parentheses).
Energies are in eV,
and the number of $d$ electrons is $N_{d} = 4.3$.}
\end{table}

\clearpage

\begin{table}
\begin{center}
\begin{tabular}{c  c c c c c }
\hline
\hline
$P$~(GPa) &
$T_{\rm m}$~(K)  &   $V_{l}$~(\AA$^3$)  &
$V_{s}$~(\AA$^3$)  &
$\Delta V / V_{s}$ (\%)  &
$\Delta S / k_{\rm B}$  \\
\hline
 $7.5   $    &    $800(100)   $   &   $      $   &  $    $  &   $    $  &  $   $  \\
 $58.7   $    &   $3100(100)  $   &   $20.74 $   &  $20.39  $  &   $1.73(5)  $  &  $ 0.73(2)  $  \\
 $88.6   $    &   $3985(100)  $   &   $18.45 $   &  $18.17  $  &   $1.54(5)  $  &  $ 0.74(2)  $  \\
 $141.1  $    &   $5200(100)  $   &   $16.09 $   &  $15.86  $  &   $1.49(5)  $  &  $ 0.74(2)  $  \\
 $204.5  $    &   $6400(100)  $   &   $14.38 $   &  $14.20  $  &   $1.28(5)  $  &  $ 0.73(2)  $  \\
 $269.0  $    &   $7450(100)  $   &   $13.23 $   &  $13.06  $  &   $1.32(5)  $  &  $ 0.75(2)  $  \\
 $333.5  $    &   $8450(100)  $   &   $12.36 $   &  $12.21  $  &   $1.20(5)  $  &  $ 0.74(2)  $  \\
 $409.5  $    &   $9450(100)  $   &   $11.57 $   &  $11.45  $  &   $1.07(5)  $  &  $ 0.74(2)  $  \\
 $491.5  $    &   $10450(100) $   &   $    $   &  $    $  &   $    $  &  $   $  \\
\hline
\hline
\end{tabular}
\end{center}
\caption{Melting temperature $T_{\rm m}$ as a function of
pressure $P$, volumes per atom $V_l$ and $V_s$ in coexisting liquid
and solid, relative volume change $\Delta V / V_s$, and entropy
of fusion $\Delta S$ of the pure exponential system for coexisting
bcc solid and liquid. Estimated errors are given in parentheses.
}
\end{table}

\clearpage

\begin{table}
\begin{center}
\begin{tabular}{c c c c c c }
\hline
\hline
$P_{\rm m}$ (GPa) & $T$ (K) & $V_{l}$ (\AA$^3$) &
$V_{s}$ (\AA$^3$) & $\Delta V / V_{s}$ (\%) &  $\Delta S / k_{\rm B}$  \\
\hline
 $(9.91)           $    &   $  2000  $   &   $  (16.99)   $   &  $ (16.27) $  &   $ (4.37) $  &  $ (1.85) $  \\
 $5.63 (50.94)     $    &   $  3000  $   &   $ 16.64 (14.53) $   &  $ 15.86 (14.23)  $  &   $  4.92 (2.93)  $  &  $ 2.75 (0.96)  $  \\
 $56.37 (111.44)   $    &   $  4000  $   &   $ 14.69 (12.82) $   &  $ 14.10 (12.55)  $  &   $  4.18 (2.15)  $  &  $ 2.19 (1.07)  $  \\
 $111.51 (163.65)  $    &   $  5000  $   &   $ 13.04 (11.91) $   &  $ 12.65 (11.71)  $  &   $  3.13 (1.65)  $  &  $ 1.53 (0.86)  $  \\
 $166.36 (215.93)  $    &   $  6000  $   &   $ 12.11 (11.27) $   &  $ 11.75 (11.10)  $  &   $  3.12 (1.56)  $  &  $ 1.46 (0.84)  $  \\
 $211.36 (281.88)  $    &   $  7000  $   &   $ 11.51 (10.64) $   &  $ 11.24 (10.49)  $  &   $  2.39 (1.46)  $  &  $ 1.09 (0.81)  $  \\
 $276.42 (373.96)  $    &   $  8000  $   &   $ 10.90 (9.95)  $   &  $ 10.61 (9.81)   $  &   $  2.76 (1.45)  $  &  $ 1.22 (0.82)  $  \\
 $335.63 (459.45)  $    &   $  9000  $   &   $ 10.38 (9.47)  $   &  $ 10.16 (9.33)   $  &   $  2.13 (1.53)  $  &  $ 0.92 (0.88)  $  \\
\hline
\hline
\end{tabular}
\end{center}
\caption{
Melting pressure $P_{\rm m}$ as a function of temperature $T$,
volumes per atom $V_l$ and $V_s$ of coexisting liquid and solid,
relative volume of fusion $\Delta V / V_s$, and entropy
of fusion $\Delta S$, for TB model at $d$-band fillings $N_d = 4.3$~(5.0).
}
\end{table}

\clearpage

{\Huge {\bf Figure Caption List}}

\begin{figure}
\centering
{\includegraphics[width=0.65\linewidth]{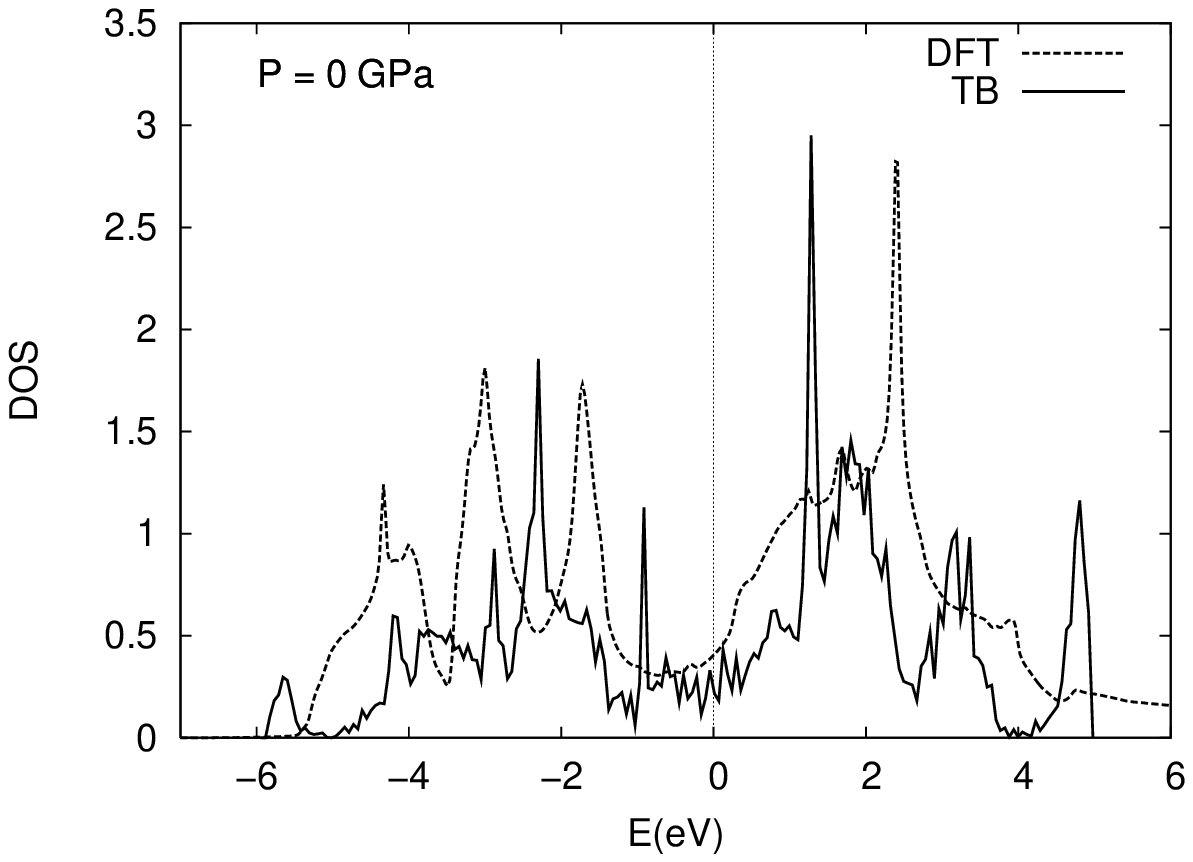} }%
{\includegraphics[width=0.65\linewidth]{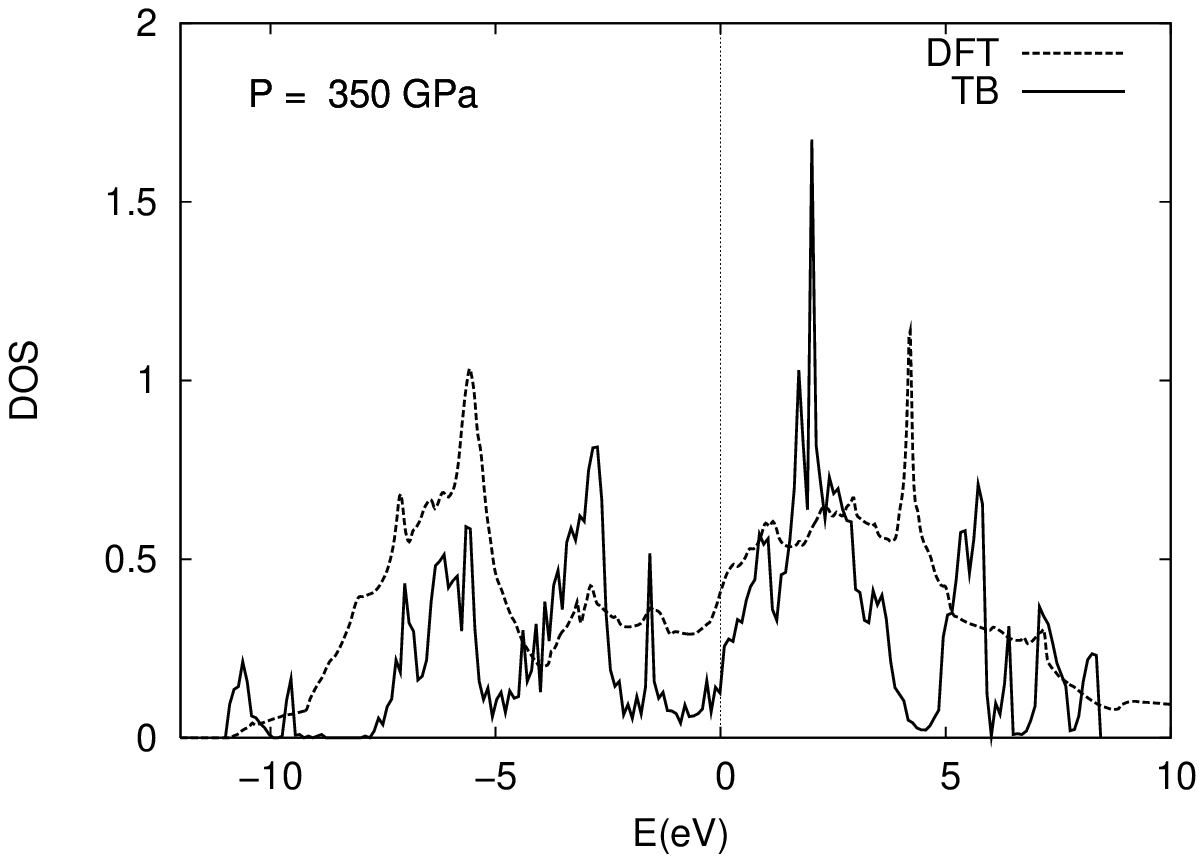} }%
\caption{}
\label{fig:dos0T}
\end{figure}

FIG.~\ref{fig:dos0T}~: $d$-component of the electronic density of states of bcc Mo
calculated at $T = 0$~K using DFT and TB  at pressures of $0$~GPa (top panel)
and $350$~GPa (bottom panel). The Fermi energy is set to zero (vertical lines).

\begin{figure}
\centerline{
\includegraphics[width=0.8\linewidth]{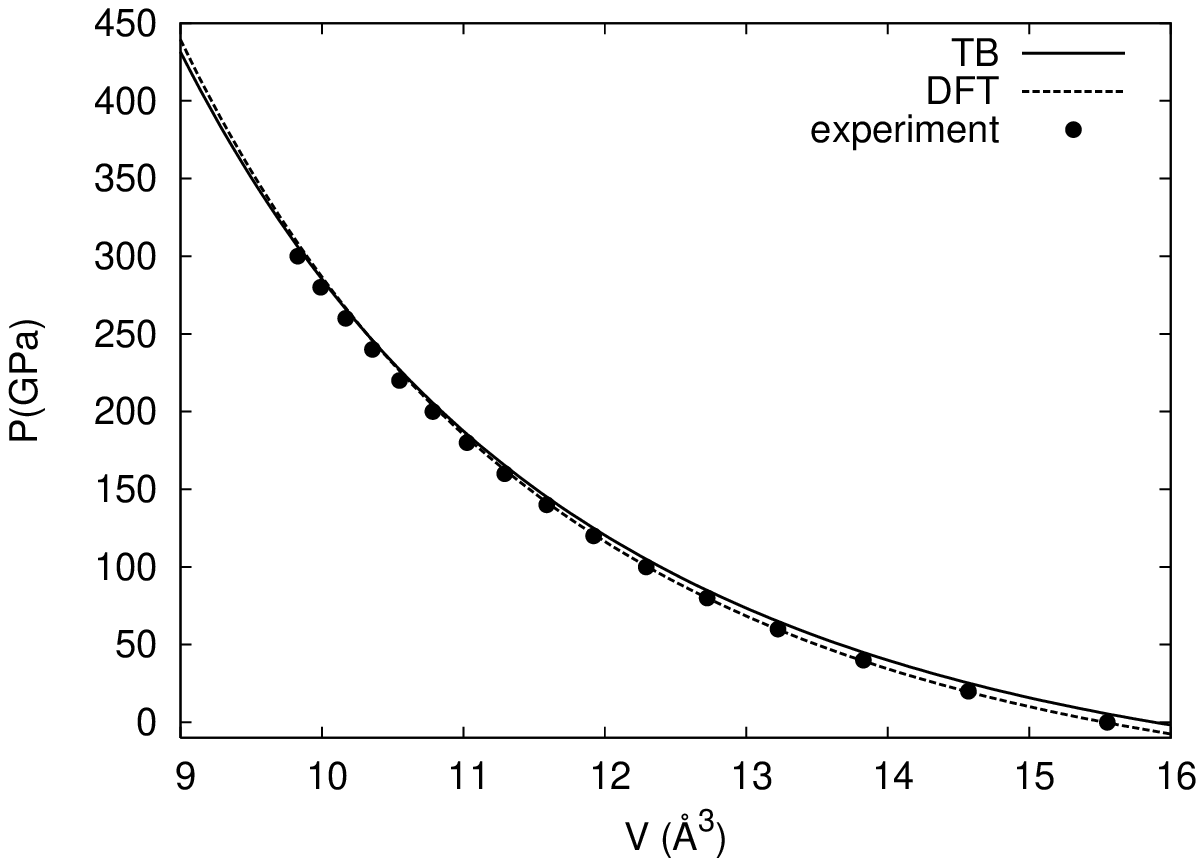}}%
\caption{}
\label{fig:eos}
\end{figure}

FIG.~\ref{fig:eos}~: Equation of state of bcc Mo obtained from
the present TB model (solid line) and  DFT (dashed line); experimental data (dots) 
from Ref.~[\onlinecite{hixson92}] are shown for comparison

\begin{figure}
\centerline{
\includegraphics[width=0.8\linewidth]{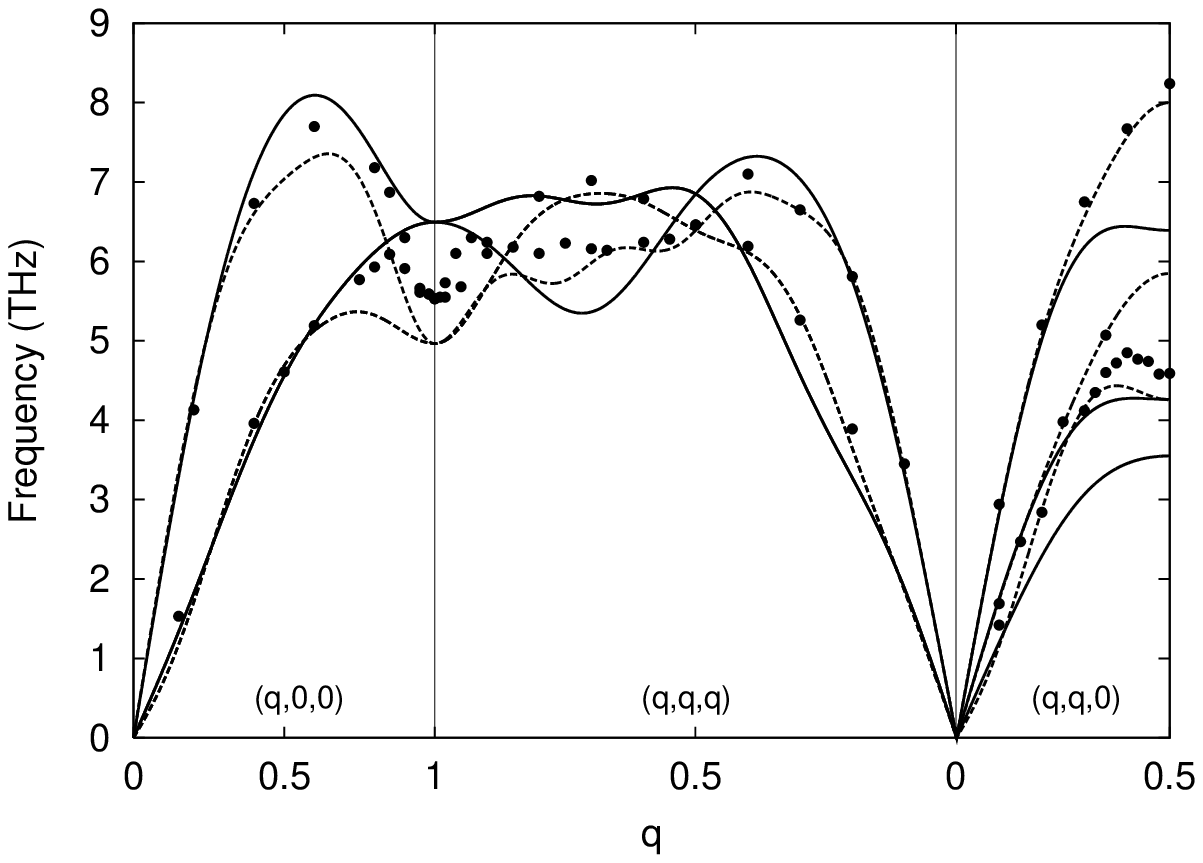}}%
\caption{}
\label{fig:phonons}
\end{figure}

FIG.~\ref{fig:phonons}~: Phonon dispersion relations of bcc Mo calculated with
the present tight-binding model (solid lines) and
DFT (dashed lines) at the experimental equilibrium volume
$V_{0} = 15.55$~\AA$^{-3}$~. Experimental data (dots) from Ref.~[\onlinecite{zarestky83}] 
are shown for comparison.

\begin{figure}
\centering
{\includegraphics[width=0.65\linewidth]{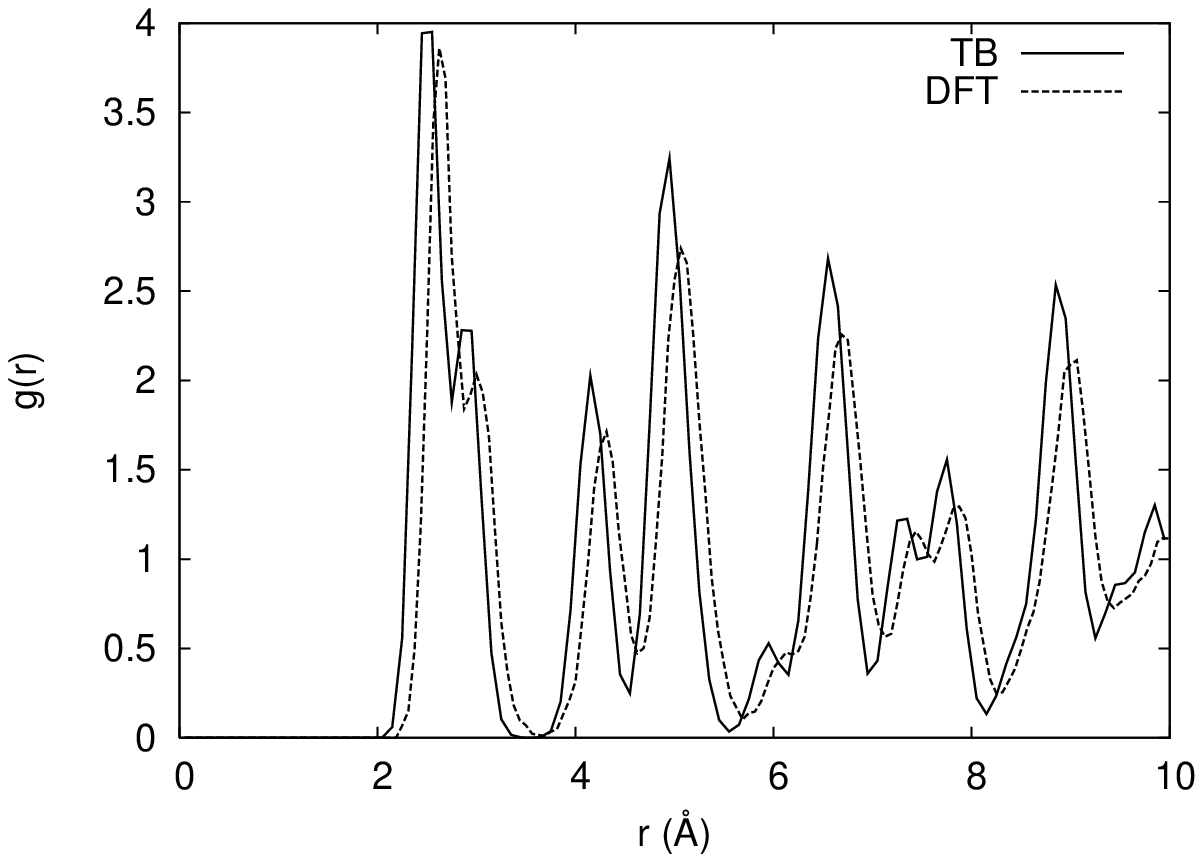} }%
{\includegraphics[width=0.65\linewidth]{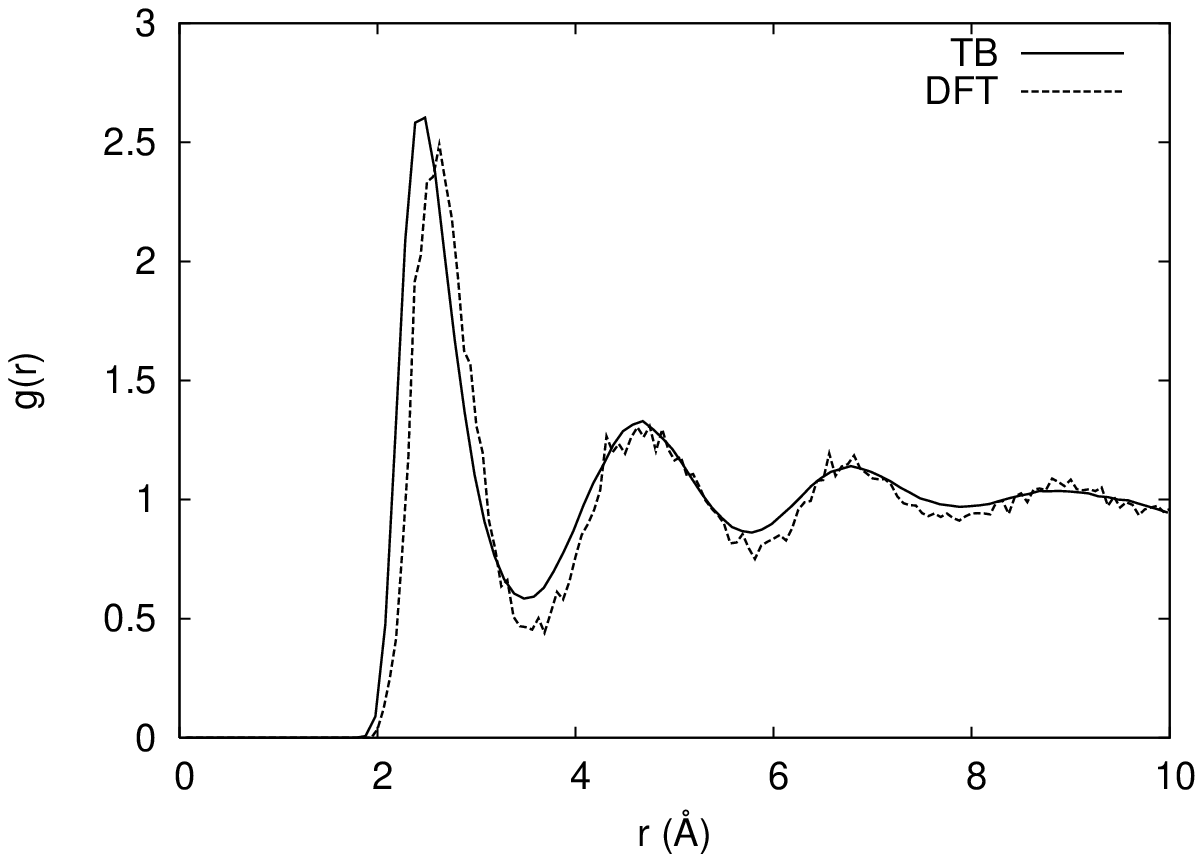} }%
\caption{}
\label{fig:grtest}
\end{figure}

FIG.~\ref{fig:grtest}~: Radial distribution function of solid
bcc Mo at $T = 2000$~K and $P = 50$~GPa from long DFT and TB m.d. runs.
\emph{Bottom}: Radial distribution function of liquid Mo at $T = 8250$~K and
$P = 250$~GPa obtained from long DFT and TB m.d. runs.

\begin{figure}
\centering
{\includegraphics[width=0.65\linewidth]{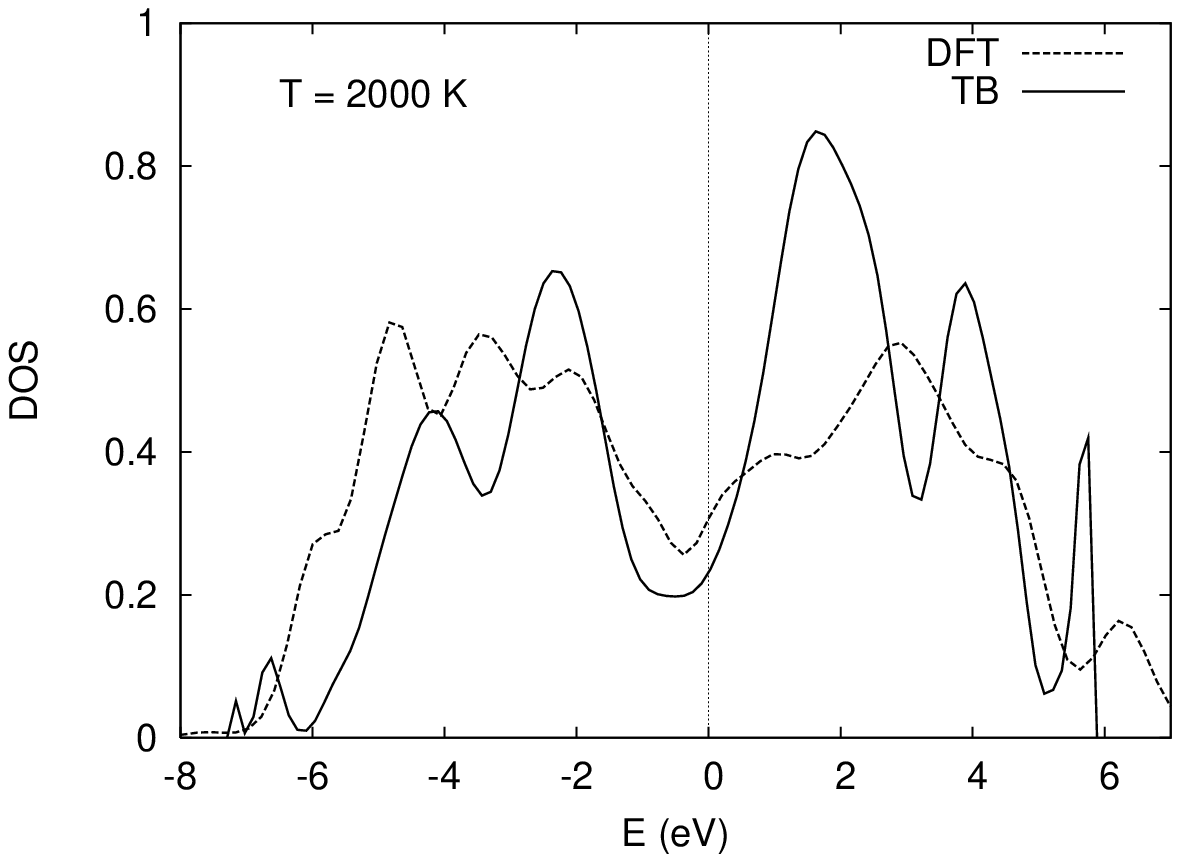} }%
{\includegraphics[width=0.65\linewidth]{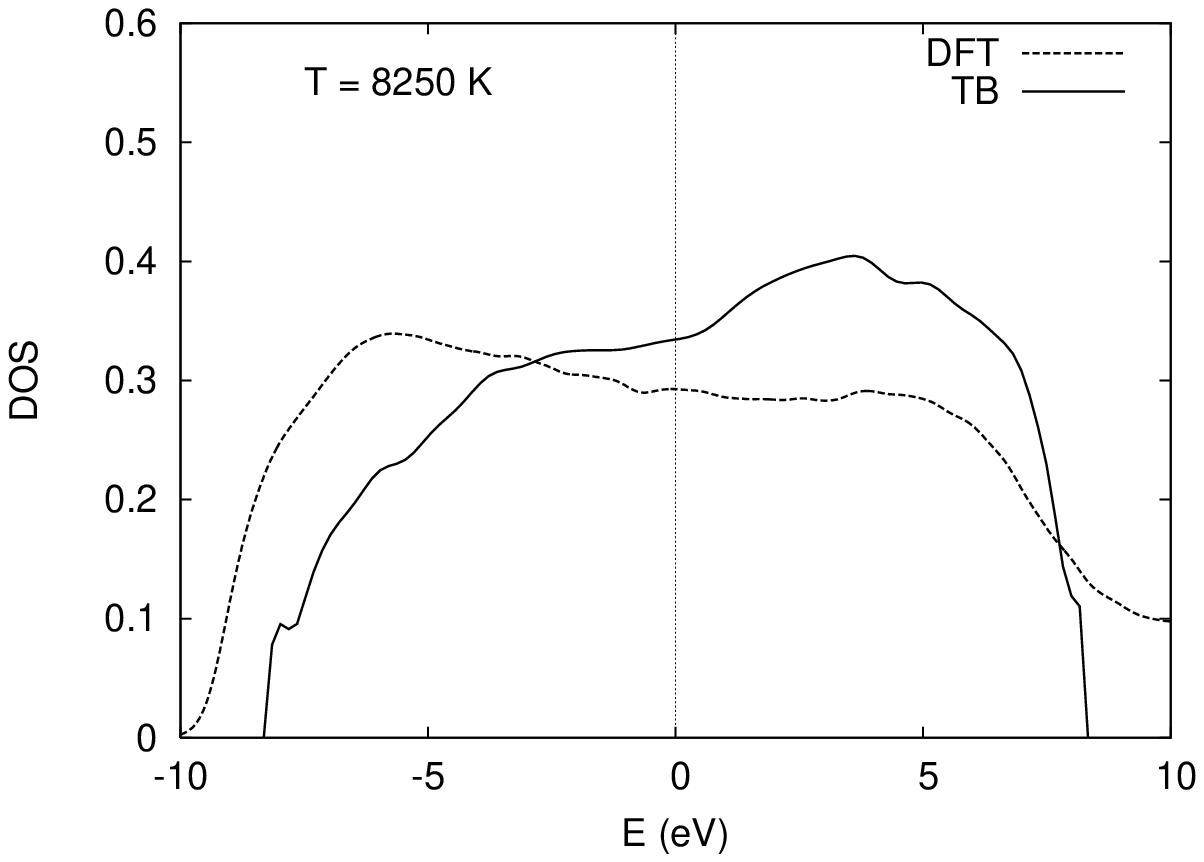} }%
\caption{}
\label{fig:dostest}
\end{figure}

FIG.~\ref{fig:dostest}~: $d$-band electronic density of states calculated by DFT 
and TB m.d. simulation for bcc Mo at $P = 50$~GPa, $T = 2000$~K (top panel)
and for liquid Mo at $P = 250$~GPa, $T = 8250$~K (bottom panel).

\begin{figure}
\centerline{
\includegraphics[width=0.8\linewidth]{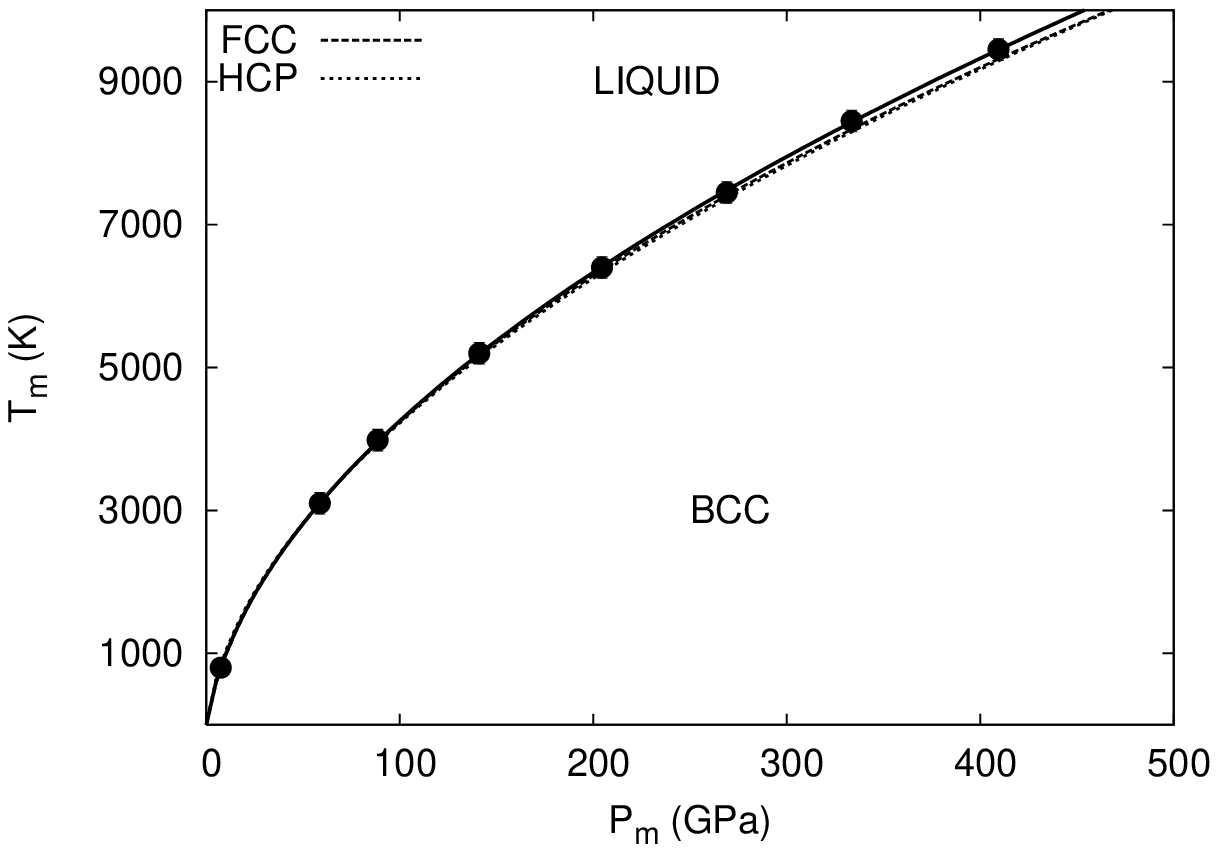}}%
\caption{}
\label{fig:meltingexp}
\end{figure}

FIG.~\ref{fig:meltingexp} Phase diagram of the pure exponential model $V_{\rm REP}$
obtained from coexisting solid and liquid phase ( $N$, $V$, $E$ )
simulations. The solid line in the figure corresponds to the bcc-liquid phase boundary
while the dashed and dotted lines are the fcc-liquid and hcp-liquid ones, respectively.
Dots simbolize points obtained directly from the phase coexistence simulations.

\begin{figure}
\centerline{
\includegraphics[width=0.8\linewidth]{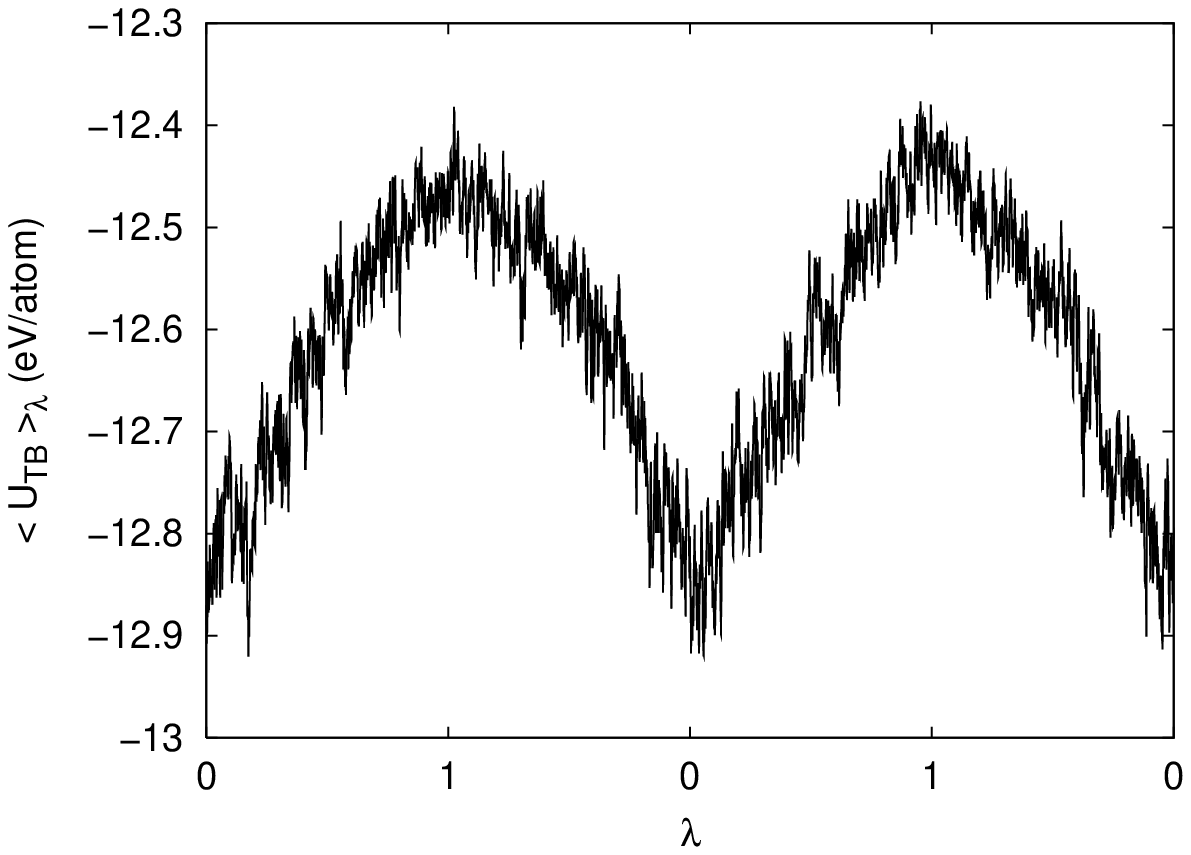}}%
\caption{}
\label{fig:adiabatic}
\end{figure}

FIG.~\ref{fig:adiabatic}~: Thermal average $\langle U_{\rm TB} \rangle_\lambda$ of
the tight-binding energy $U_{\rm TB}$ as function of $\lambda$
in an adiabatic thermodynamic-integration calculation of the free
energy difference $F_{\rm tot} - F_{\rm REP}$ between the
REP+TB and REP systems. The plot shows
$\langle U_{\rm TB} \rangle_\lambda$ from a simulation in which
$\lambda$ executes a double cycle
$0 \rightarrow 1 \rightarrow 0 \rightarrow 1 \rightarrow 0$,
the rate of variation $| d \lambda / d t |$ being
$1/9$~ps$^{-1}$.

\begin{figure}
\centerline{
\includegraphics[width=0.8\linewidth]{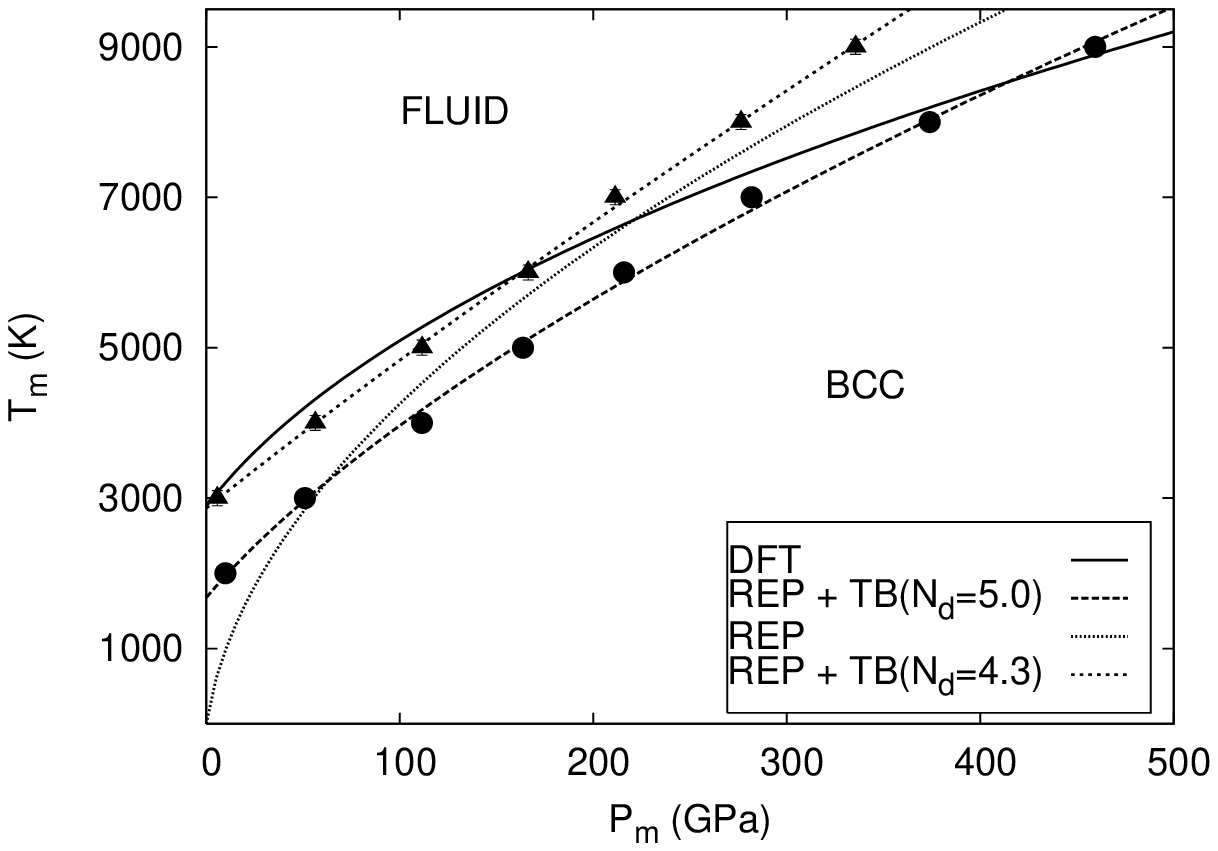}}%
\caption{}
\label{fig:melting}
\end{figure}

FIG.~\ref{fig:melting}~: Melting curve of TB model at
$d$-band fillings $N_{d} = 4.3$ (dashed line) and $5.0$ (dotted line).
The melting curve of the pure exponential model and that of Mo from DFT
simulations~[\onlinecite{cazorla07}] are show for comparison.

\begin{figure}
\centerline{
\includegraphics[width=0.8\linewidth]{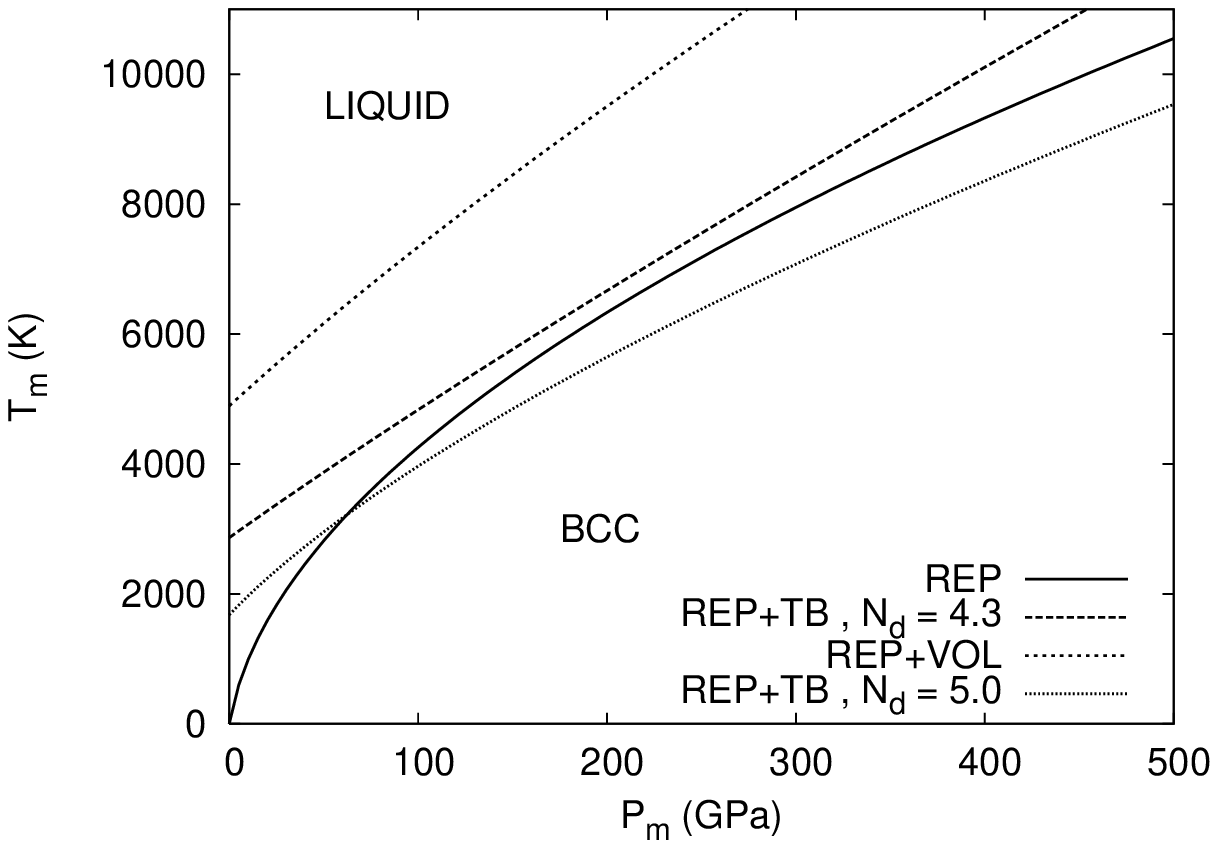}}%
\caption{}
\label{fig:meltingexpvol}
\end{figure}

FIG.~\ref{fig:meltingexpvol}~: Melting curve of the repulsive pure exponential
potential $V_{\rm REP}$ (solid line), pure exponential potential plus a bonding 
energy term depending just on volume $V_{\rm REP} + E_{d}$ (short-dashed line), 
and full tight-binding model $U_{\rm tot} = V_{\rm REP} + U_{\rm TB}$ at 
$N_{d} = 4.3$ (long-dashed line) and $5.0$ (dotted line)~.

\begin{figure}
\centering
{\includegraphics[width=0.65\linewidth]{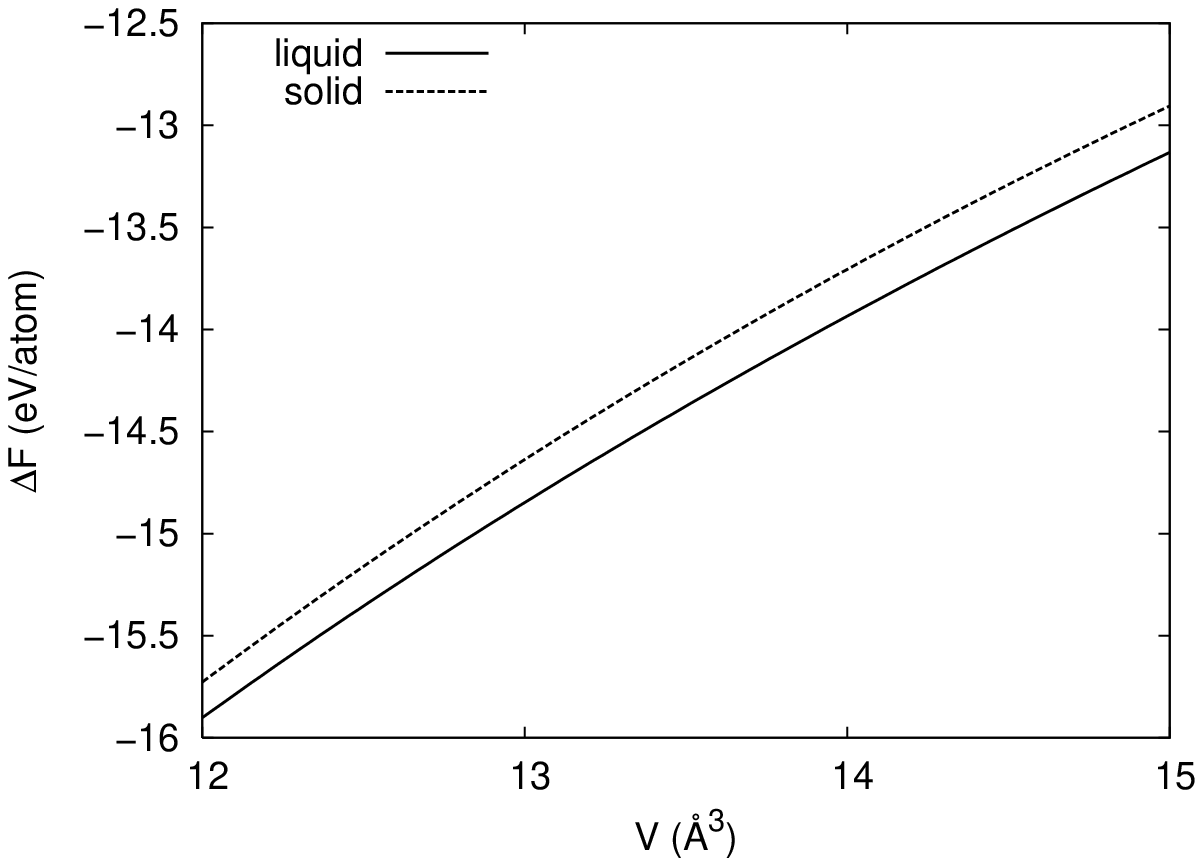} }%
{\includegraphics[width=0.65\linewidth]{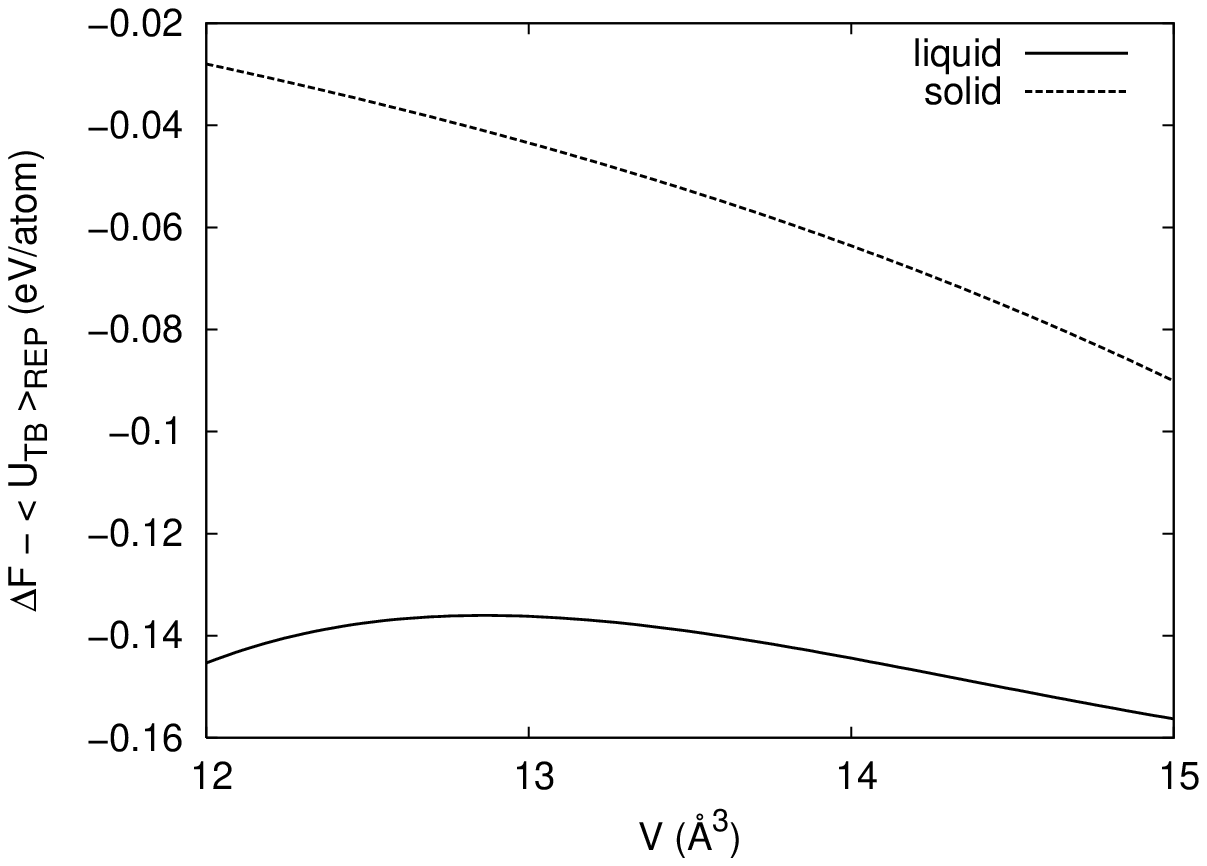} }%
\caption{}
\label{fig:deltaf}
\end{figure}

FIG.~\ref{fig:deltaf}~: \emph{Top}: Free energy difference
$\Delta F = F_{\rm tot} - F_{\rm REP}$ of the total
tight-binding model and repulsive pure exponential potential
in the liquid and solid
phases at temperature $ T = 6000$~K and for $N_{d} = 4.3$~.
\emph{Bottom}: Quantity $\Delta F - \langle U_{\rm TB} \rangle_{\rm REP}$
in the liquid and solid phases at temperature
$ T = 6000$~K and for $N_{d} = 4.3$~.

\clearpage

\clearpage

\clearpage

\clearpage

\clearpage

\clearpage

\clearpage

\clearpage

\clearpage

\clearpage

\end{document}